\documentclass[11pt]{article}
\newcommand{\be}{\begin{equation}}
\newcommand{\ee}{\end{equation}}
\newcommand{\beq}{\begin{eqnarray}}
\newcommand{\eeq}{\end{eqnarray}}

\usepackage{graphicx}

\begin{document}
\begin{center}
{\bf\LARGE The Birth and Death of a Universe }\\
[7mm]
\vspace{1cm}  
H. M. FRIED
\\
{\em Department of Physics \\
Brown University \\
Providence R.I. 02912 USA}\\
fried@het.brown.edu\\
[5mm]
Y. GABELLINI
\\
{\em Institut Non Lin\'eaire de Nice\\
UMR 7335 CNRS\\
 1361 Route des Lucioles\\
06560 Valbonne France}\\
yves.gabellini@unice.fr\\
[5mm]

\vspace{4cm}
Abstract
\end{center}
This letter is meant to be a brief survey of several recent publications providing a simple, sequential explanation of Dark Energy, Inflation and Dark Matter, which leads to a simple picture of the why and the how of the Big Bang, and thence to a possible understanding of the birth and death of a Universe. 
\newpage
This review begins by noting a new, QED--based derivation of dark energy that is able to provide the amount of quantum vacuum energy density which Astrophysicists believe is responsible for the continuing expansion of our universe \cite{FG1}. Fig.1 represents a picture of the shape of our energy density. 

\begin{figure}
\includegraphics[width=10truecm]{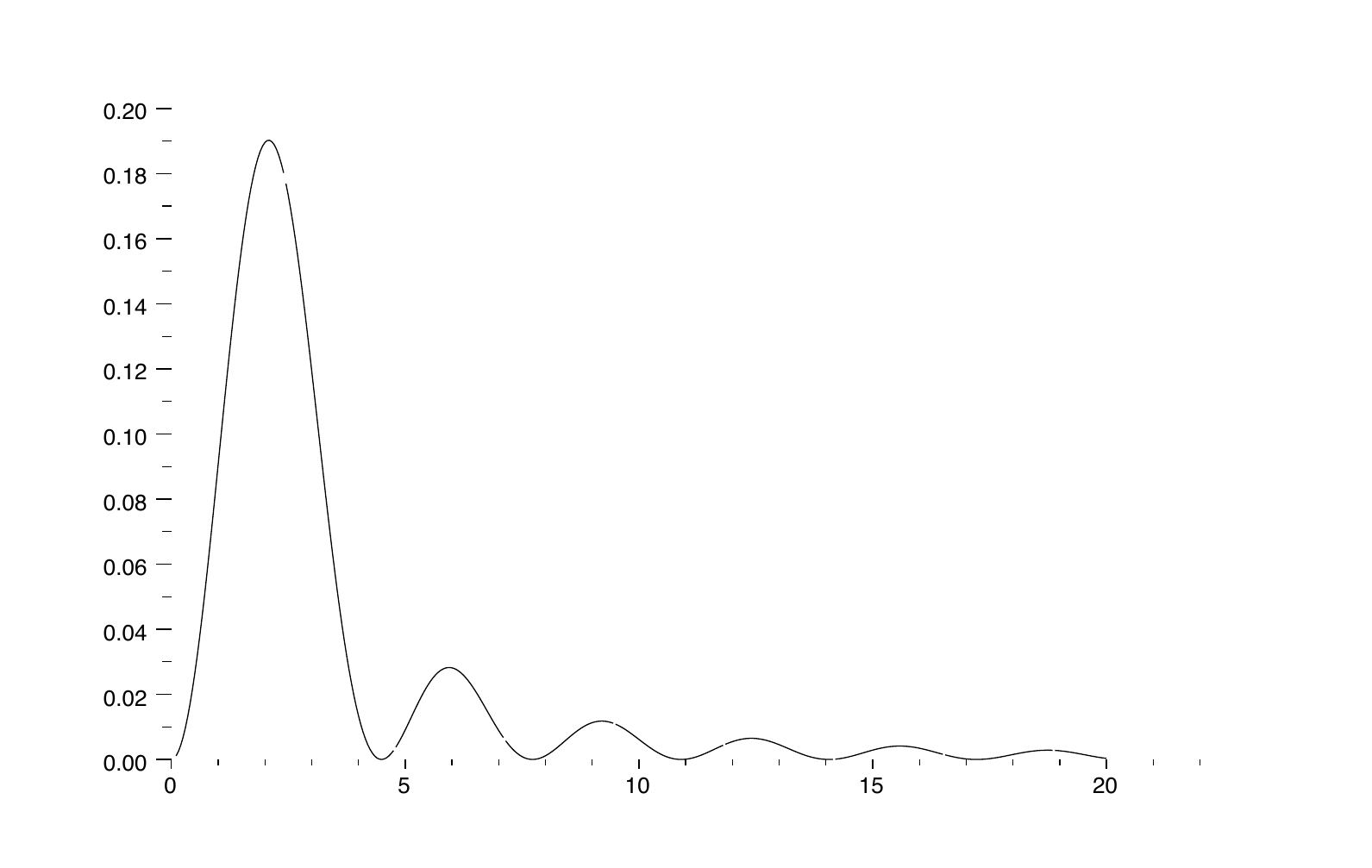}
\caption{A plot of $f(x) = \displaystyle{1\over x^2}\Bigl(\cos x - {\sin x\over x}\Bigr)^2$ vs. $x$}
\end{figure}

We have then found a surprisingly simple but unexpected extension of that dark energy analysis to account for inflation, the violent explosion of matter and energy from the Big Bang, representing the very beginning and later components of our universe \cite{FG2}.

	That extension was made possible by the necessary assumption of the existence of electrically charged, fermionic pairs of tachyons ($T$) and anti-tachyons ($\bar T$) fluctuating in the quantum vacuum, in addition to corresponding lepton -- and quark -- pair fluctuations. And the mass of such tachyons can be chosen to be slightly less than, on the order of, or even greater than the Planck mass, $M_P$, which provide fine fits to the initial and final times of inflation, and its corresponding energy density values, as listed in the cosmological tables for inflation \cite{LL}\cite{CP}.

	A further question arises concerning the possibility of the extraction of such $T-\bar T$ pairs from the quantum vacuum into the real vacuum of the everyday world. This event would surely be possible in a region close to the explosion of a Super Nova, where one might expect huge electric fields to be present for a short time, which would have the effect of tearing such $T-\bar T$ pairs apart, and hurling them, in opposite directions, into the real vacuum as an example of an extremely energetic Schwinger mechanism. From the kinematics and Born approximation analyses of photon emission and absorption from and by an electrically charged fermionic tachyon \cite{FG3}, it is easy to see that a very high energy, charged tachyon could be a perfect example of a dark matter particle, which reabsorbs almost every photon that it emits. 
	
	The derivation of dark energy, inflation, and dark matter qualitatively described in the above paragraphs, then leads to a possible understanding of the why and the how of the Big Bang, as well as to the title of these paragraphs. For this, one must understand the current cosmological belief that an energy density, $\rho$, proportional to  $M_P ^4$, is the largest value of such a density which our equations containing Quantum Mechanics, Relativity, and Gravity can support; were there to suddenly appear an energy density of value larger than that, then in the region of that $\rho$ there must follow a breakdown of at least one of our standard descriptions of those three theories above.

	With this in mind, consider a  very high energy, charged, fermionic  tachyon propagating in remote galactic space, which particle has received a large percentage of its high energy by absorbing CMB photons in the millions of years that it has been in motion. Suddenly, and unexpectedly, it meets a $\bar T$ of the same species; and they annihilate. At that tiny spot of annihilation, the energy density will be larger than the $M_P^4$ of the previous paragraph, and something has to give. Let us suppose that what immediately happens is that -- at that spot of annihilation -- the separation between the quantum vacuum and the real vacuum is disrupted, and the huge energies contained in the quantum vacuum can suddenly explode into the real vacuum. This is the Big Bang of a new universe, with origin at that point of annihilation, in which energy, and charge contained in the quantuum vacuum of the old universe are blasted into the new universe, which has no memory of its origin. Some of that energy goes into the quantum vacuum of the new, and growing universe, while the remainder forms the mass of a part of this new universe. 

	And what is the fate of the old universe ? Consider a portion of its matter, an ordinary galaxy, which is located far from the origin of the new universe. What has kept that galaxy in existence for all of its previous lifetime is the energy stored in that part of the total quantum vacuum in its immediate vicinity, for it is that portion of the quantum vacuum which fights against the natural, gravitational attraction of each segment of its matter to collapse into each other. But now, as compressed gas in a punctured balloon far from the puncture point tends to move towards that point of escape, so can the quantum vacuum energy near that galaxy tend to move away, towards the point of annihilation, leaving the galaxy with only a fraction of its defense against gravity, a fraction which gets smaller and smaller as time goes by. 

	And what is the fate of that old galaxy ? It takes little imagination to realize that it will be squeezed out of existence, forming a supermassive black hole; that is, a black hole containing the mass of perhaps a million of its previous suns. Eventually, that supermassive black hole will become the center of another galaxy of the new universe; and on one of its planets, astronomers will one day discover -- as have ours, within the last few years -- that the black hole at the center of their galaxy -- and of so many other galaxies -- is supermassive.

	As a summary, our speculations provide a sequential and coherent picture of the three, modern cosmological mysteries -- Inflation, Dark Matter, and Dark Energy -- and of the origin and possible fate of our Universe.

\vskip1truecm {\bf\Large Acknowledgement}
\vskip0.3truecm 
It is a pleasure to acknowledge helpful conversations with I. Dell'Antonio.

This publication was made possible through the support of the
Julian Schwinger Foundation.


\vskip1.5cm

\begin{thebibliography}{**}


\bibitem{FG1}  H.M. Fried, Y. Gabellini, QED vacuum loops and vacuum energy, Eur. Phys. J. C (2013) 73 : 2642.

\bibitem{FG2} H.M. Fried, Y. Gabellini, QED vacuum loops and inflation, Eur. Phys. J. C (2015) 75~ : 125

\bibitem{LL} A.R. Liddle and D.H. Lyth, Cosmological Inflation and Large Scale Structure, Cambridge University Press (UK) 2000.

\bibitem{CP} Planck 2013 results. XVI. Cosmological parameters, arxiv.org/abs/1303.5076.

\bibitem{FG3} H.M. Fried and Y. Gabellini, Quantum Tachyon Dynamics, arXiv-hep/th 0709.0414 (2007). 


\end{thebibliography}
\end{document}